\documentstyle{black}
\def\ga{\gamma}
\def\etall{{\it et al., }}
\def\etal{{\it et al.}}
\begin{document}

\title{%
GAMMA-RAY BURSTS - THE SECOND REVOLUTION}

\author{Tsvi PIRAN \\
{\it Racah Institute for  Physics, The Hebrew  University,}\\ {\it Jerusalem, 91904 Israel,
tsvi@nikki.fiz.huji.ac.il}}

\maketitle

\section*{Abstract}
Gamma-ray bursts GRBs are among the most mysterious astronomical
phenomenon ever discovered. Unlike most astronomical discoveries which
were explained within weeks or months after their initial discovery,
GRBs remain a puzzle for more than thirty years.  During the last
decade our understanding of GRBs has undergone two major revolutions.
First, BATSE discovered that GRBs are distributed isotropically over
the sky and thereby demonstrated their cosmological origin. The second
revolution tool place more recently when BeppoSAX discovered GRB
afterglow. This confirmed the fireball model and led to a wealth of
observational data, some of which has not been fully understood yet.
The emerging picture is that GRBs are the most luminous objects and
the most relativistic objects ever discovered: (i) GRBs involve
relativistic motion at a velocity of 0.9999c or larger.  (ii) Most
current GRB models involve the formation of a black hole in one way or
another.  (iii) If binary neutron star mergers are the sources of GRBs
then GRBs are also associated with gravitational radiation signals.
Finally, (iv) as cosmological power-houses that are observed to high
red-shift GRBs might be used to measure cosmological parameters and to
teach us about the epoch of galaxy formation.

\section{Introduction}

Our understanding of GRBs has undergone two major revolutions during
the last decade. The first revolution took place at the early nineties
when BATSE discovered that the angular distribution of GRBs is
isotropic and there is a paucity of weak bursts (Meegan 1992). This
discovery ruled out the then popular galactic disk neutron star model
and established that GRBs are cosmological with a typical red-shift of
order unity (Piran 1992; Fenimore \etal 1993; Cohen \& Piran
1995). The immediate implication was that the total energy involved is
of order $10^{52}$ ergs or more. Since this energy is released within
a few seconds GRBs are the most (electromagnetically) luminous objects
in the Universe.

The rapid variability of GRBs suggests a compact source.  When
combined with the observed flux and the cosmological distance this implies
enormous photon density. Such a source will have an huge optical depth
for pair productions (by the high energy $\gamma$-rays).  However, the
observed spectrum extends far above 500KeV and the spectrum is non
thermal - indicating that the sources are optically thin.

A solution for the problem was suggested even before BATSE's
discoveries.  The compactness problem could be resolved if the
emitting regions are moving relativistically with a Lorentz factor
$\gamma \ge 100$. This have led to the fireball model. According to
this model GRBs are produced when a relativistic flow is slowed down
and its kinetic energy is extracted and converted to radiation. This
model leaves open the questions what is the ``inner engine'' - the
source that produces the relativistic flow and how is the flow
produced?  It deals only with the emitting regions, which are distant
and separate from the source.

An immediate prediction (Paczy\'nski \& Rhoads 1993; Katz 1994;
M\'esz\'aros \& Rees 1997; Sari \& Piran 1997) of the fireball model
is the appearance of afterglow - a following radiation that results
from stages when the flow is slower, but still relativistic. The
afterglow is produced on much longer time scale and at longer
wavelengths. The second GRB revolution occurred on 28 February 1997
when the Italian-Dutch satellite BeppoSAX discovered X-ray emission
that accompanied GRB970228 (Costa \etal 1997). The X-ray emission
continued for a few hours.  The accurate X-ray position enabled follow
up observations that detected a decaying optical source.  GRB
afterglow was discovered and an optical and X-ray counterparts were
found. Since then about a dozen afterglows that accompanied GRBs were
observed. Cosmological redshifts were measured in several cases,
confirming the cosmological origin of GRBs. Evidence for relativistic
motion and relativistic shocks was found and the fireball model was
practically confirmed.

In this talk I review the fireball model. In particular I examine the
{\it external-internal shock model}, according to which  GRBs are
produced by internal  shocks within the flow while the afterglow is
produced by external shocks with the ISM. I will show how the recent
observations support this picture.  However, as commonly happens when
new data becomes available new puzzles arise. I will briefly discuss
the new puzzles that arose in recent observations.

As the focus of this conference, honoring Humitaka Sato, is on general
relativity it is worthwhile mentioning that GRBs are not only the most
luminous objects known, they are also the most relativistic objects
ever discovered.  First, the generic fireball model is based on a
relativistic macroscopic motion at a velocity of 0.9999c.  This is
closer to the speed of light than what was observed in any other
macroscopic object. The GRB and the associated fireball involve
relativistic shocks which are observed for the first time.  Most
models for the source that produces the fireball involve compact
objects and end up in a black hole. Thus GRBs signal, most likely, the
formation of black holes.  The binary neutron star merge is currently
the leading GRB model.  In this case GRBs should appear in coincidence
with the unique gravitational radiation signals produced by these
mergers.  Finally, GRBs are a cosmological population that is observed
regularly and uniformly.  Various cosmological relativistic effects
such as redshift and time dilation have already been observed in the
GRB data.  When understood, GRBs could teach us a lot about the epoch
of galaxy formation and possibly could provide new ways to measure
cosmological parameters.

\section{GRB Observations}

GRBs are short and intense pulses of low energy gamma-rays. The bursts
duration, $T$, varies from a fraction of a second to several hundred
seconds.  Within this time scale most bursts are highly variable and
the variability time scale $\delta T \ll T$. A typical ratio is $
{\delta T/ T} \sim 0.01$.  The spectrum is non thermal. It is fairly
well approximated by the Band Spectrum - two power laws joint smoothly
together.  GRBs contain a significant power in the higher parts of the
spectrum and GeV emission has been observed in many bursts. The
nonthermal spectrum indicates that the radiation emerges from
optically thin regions.  The slop of the lower part of the spectrum
indicates that GRBs are produced by synchrotron emission in
relativistic shocks (Cohen \etal 1997).  GRBs appear from random
positions in the sky with no repetition.  The distribution of GRBs on
the sky is isotropic, suggesting their cosmological origin. The
paucity of weak bursts is consistent with a cosmological origin as
well. In fact one can infer from the peak flux distribution that BATSE
detects a typical burst from a redshift of $z \sim 1$ (see e.g. Piran
1992; Fenimore \etal 1993; Cohen \& Piran 1995). The agreement of the
peak flux distribution with a cosmological standard candles
distribution suggests that the luminosity function of GRBs is not too
broad and that GRB luminosities do not vary by more than one order of
magnitude (Cohen \& Piran 1995).

\section{The Fireball Model}

The fireball model is a generic GRB model according to which GRBs form
when a relativistic expanding shell (or a relativistic jet) is slowed
down and its energy is converted to gamma-rays.  This model was
motivated by the need to overcome the compactness problem. It is the
starting point of our discussion. There are two variants of the
fireball model, the {\it external shock model} and the {\it internal
shock model}. I will show that both shocks take place in an {\it
external-internal shock model}: the GRBs are produced by internal
shocks while the afterglow is produce by an external shock that
follows.

\subsection{Compactness and Relativistic Motion}

The key to understanding GRBs lies, I believe, in understanding how
GRBs bypass the compactness problem.  Consider a typical burst with a
total energy of $10^{52}$ergs (as inferred from the observed flux and
the implied distance of a cosmological source) that varies on a time
scale $\delta T \approx 10$msec.  Standard considerations suggest that
the temporal variability implies that the sources are compact with a
size, $R_i < c \delta T \approx 3000$km.  The resulting energy density
and photon density at the source are enormous. Under these conditions
photons with $h \nu \ge m_e c^2$ would interact with lower energy
photons ($h \nu' \ge (m_e c^2)^2 /h \nu$) and produce electron-positron
pairs.  The typical optical depth for this process is $\sim
10^{16}(E/10^{52}{\rm ergs}) (\delta T / 10~{\rm msec})^{-2}$
(Piran 1997). However, the observed non-thermal spectrum contains
many high energy photons which indicate with certainty that the
source must be optically thin.

The compactness problem can be resolved if the emitting region is
moving towards us with a relativistic velocity characterized by a
Lorentz factor, $\gamma \gg 100 $.   The photons' energy
at the source would be  lower by a factor $\gamma$ than the observed energy.
This implies that fewer photons have sufficient energy to produce
pairs.  Additionally, relativistic effects allow the radius from which
the radiation is emitted to be larger than the previous estimate by a
factor of $\ga^2$: $R_e \le \gamma^2 c \delta T$.  The resulting
optical depth is lower by a factor $ \ga^{(4+2\alpha)}$ (where $\alpha
\sim 2$ is the spectral index). The source will be  optically thin
if it is moving towards us with a Lorentz factors $\ga >
10^{16/(4+2\alpha)} \approx 10^2$.

Relativistic motion does not necessarily mean that the center of mass
of the source is moving towards us at such a high velocity. In fact
energetic considerations suggest that this is highly unlikely. Instead
the motion towards us (the observers) could occur most naturally from
a spherical shell that is expanding relativistically.  A
relativistically expanding spherical shell would be observed from all
directions. Relativistic beaming will take place here in the sense
that each observer will see only a small portion of the shell.  It
will see radiation only from a region $\ga^{-1}$ away from the line of
sight to the center.  This means that when we observe a source we
cannot know whether it is actually spherical.  In fact the burst may
originate from a jet with an angle $\theta$. In this case it will be
observed within a solid angle $max[\theta^2,\ga^{-2}]$ along the jet.
If $\theta > \ga^{-1}$ we wouldn't be able to tell that we are
observing a jet.

The potential of relativistic motion to resolve the compactness
problem was realized in the eighties by Goodman (Goodman 1986),
Paczy\'nski (Paczy\'nski 1986) and Krolik \& Pier (Krolik \& Pier
1991).  While Krolik \& Pier (Krolik \& Pier 1991) considered a
kinematical solution, Goodman (Goodman 1986) and Paczy\'nski
(Paczy\'nski 1986) realized that required relativistic motion could
arise naturally when a large amount of energy is released within a
small volume. The large optical depth of the radiation leads to a
formation of a photon pairs-radiation fluid which expands
relativistically under its own pressure.  This was called a fireball.
A pure radiation fireball expands until it becomes optically thin and
then all the radiation escapes.  However, the resulting spectrum turns
out to be almost thermal (Goodman 1986) and thus a pure radiation
fireball cannot serve as a model for GRBs.

Shemi \& Piran (Shemi \& Piran 1990) and Paczy\'nski (Paczy\'nski
1990) have shown that the resulting fireball will be drastically
different if it contains even a small amount of baryonic mass. First
the electrons associated with the baryons will dominate the opacity
and the fireball will become optically thin at a latter stage.  More
important however is the dynamical results.  All the initial energy
will be given eventually to the kinetic energy of the baryons.  If the
rest mass is sufficiently small ($Mc^2 \ll E$) the fireball will
become relativistic with an asymptotic Lorentz factor $\gamma = E/M
c^2$. The baryons form a cold shell.  The width of the shell,
$\Delta$, equals the initial size of the fireball if the source
producing the fireball is ``explosive'' and it release all its energy
at once. If, on the other hand the source operates for a long time
(long compared to its light crossing time) then $\Delta = c t$.  In
this case we expect that fluctuations in the flow (which are essential
for the formation of internal shocks) would be on a length scale
$\delta$ greater or equal to the size of the source, $ R_{source}$.

Relativistic velocities were suggested as a theoretical concept as the
only way to overcome the compactness problem.  Their appearance in
GRBs was recently established observationally in afterglow
observations (Frail \etal 1997).  This is one of the great successes
of the ``second GRB revolution''.

\subsection{Relativistic Shocks}

The shell's kinetic energy could be converted to ``thermal'' energy of
relativistic particles (and then to gamma-rays via synchrotron or
inverse Compton) via shocks (M\'esz\'aros \& Rees 1992).  Both the low
energy spectrum of GRBs and the high energy spectrum of the afterglow
provide indirect evidence for relativistic shocks in the GRB (Cohen
\etal 1997) and in the afterglow (Wijers, Rees \& M\'esz\'aros 1997).
The shocks could be (i) {\it external} - due to interaction with an
external medium like the ISM (M\'esz\'aros \& Rees 1992) or (ii) {\it
internal} due to irregularities in the flow itself (Narayan,
Paczy\'nski \& Piran 1992; Rees \& M\'esz\'aros 1994; Paczy\'nski \&
Xu 1994). In either case these shocks must take place at sufficiently
large radii where the flow is optically thin, allowing the emission of
a non-thermal spectrum.

Consider a shell of width $\Delta$. The interaction of the shell with
the ISM will take place in the form of two shocks.  A forward shock
propagating into the ISM and a reverse shock propagating into the
shell. The behaviour of external shocks depends on a dimensionless
parameter $\xi \equiv (l/ \Delta )^{1/2} \ga^{-4/3}$, where $l=(3E/4
\pi n_{ism} m_p c^2)^{1/3} $ is the Sedov length, the radius of a
sphere in which the external rest mass equals the energy of the
fireball.  The shell's kinetic energy  is converted to
shocked particles via  external shocks  at the radius:
\begin{equation}
R_{ext}=\cases{ l/\gamma^{2/3} & if $\ \ \xi > 1$, \cr
l^{3/4} \Delta^{1/4} & if $ \ \ \xi < 1$.}
\end{equation}
Clearly to produce the non-thermal GRBs, this should take place in an
optically thin region.  Thus: $R_{ext} < R_\tau=\sqrt{ \sigma_T n_{ism} l^3
/\ga }$ .

The observed duration is  $T$:
\begin{equation}
T={R_{shock}\over c \ga^2_{shock}} =
\cases{ l/c \gamma^{8/3} & if $\ \ \xi > 1$, \cr
\Delta /c & if $ \ \ \xi < 1$,}
\label{time}
\end{equation}
where $\ga
_{shock}$ is the Lorentz factor of the shocked material.

An observer detects emission from up to an angle $\ga^{-1}$ from the line
of sight. Radiation from different angle arrives at different times
with a typical spread of $R_{ext}/c \gamma_{shock}^2$. This is of the
same order of magnitude as $T$ given by equation \ref{time} and
consequently external shock must be smooth and they cannot show a
variable temporal structure. Since most bursts are highly variable Sari
\& Piran (1997) concluded that GRBs cannot be produced by
external shocks. The afterglow is, on the other hand, smooth and it
can be naturally generated via the interaction of the shell with the
ISM.

Internal shocks are an alternative mechanism for conversion of the
kinetic energy.  Such shocks take place when the flow is
irregular. Let a typical irregularity have a distance scale
$\delta$. Then internal shocks will occur at:
\begin{equation}
R_{int} = \delta \ga^2 .
\end{equation}
For consistency, $R_{int}$ should be smaller than $R_{ext}$, otherwise
external shocks will take place first.  It should also take place in
the optically thin regime: $R_{int} < R_{\tau}$.  For internal shocks
$\Delta > R_{int}/\ga^2$ and the burst's duration is $T=\Delta/C$. The
burst varies on a time scale $\delta t = \delta /c = R_{int}/\ga^2 c$.
since $\delta \ll \Delta$ the variability condition $\delta t \ll T$
can be easily satisfied. However this requires $R_{source} \le \delta
\le 10^3$km.

Internal shocks can convert only a fraction (at most 40\%) of the
kinetic energy of the shell to radiation (Kobayashi, Piran \& Sari
1997; Mochkovitch, Maitia \& Marques 1995). Sari \& Piran (1997)
suggested that the rest of the energy will be emitted latter when the
shell encounters the ISM. According to this {\it internal external
shock model} the resulting radiation from this external shock will not
necessarily be in gamma-rays - it would appear as a following emission
in other parts of the electromagnetic spectrum - afterglow.  Afterglow
was predicted earlier by various authors (Paczy\'nski \& Rhoads 1993;
Katz 1994; M\'esz\'aros \& Rees 1997). However it was generally
assumed that the GRB and the afterglow are produced by different
stages of the same external shock. In this case the afterglow would
have been directly related and scaled to the GRB. However, GRB and
afterglow observations have revealed that there is no direct scaling
between the two phenomena. This fits naturally the prediction of the
{\it internal-external model} (Sari \& Piran 1997) in which the GRB
and the afterglow are produced by two different phenomena.

It should be stressed that within the fireball model the GRB and the
afterglow are produced when a relativistic ejecta is slowed
down. According to this picture the ``inner engine'' the source of the
GRB remains hidden and unseen.  No observed radiation emerges directly
from it.

\section{Afterglow  - The second Revolution}

GRB observations were revolutionized on February 28 1997 with the
discovery of an X-ray counterpart to GRB970228 by the Italian-Dutch
satellite BeppoSAX (Costa \etal 1997). X-ray observations by BeppoSAX,
ROSAT and ASCA revealed a decaying X-ray source whose flux $\propto
t^{-1.33\pm0.11}$. The accurate position determined by BeppoSAX
enabled the identification of an optical afterglow (van Paradijs \etal
1997) - a decaying point source surrounded by a red nebulae.  The
nebula's intensity does not vary, while the point source decays  as
a power law  $\propto t^{-1.2}$ (Galama 1997).  Afterglow was also
detected from GRB970508.  Variable emission in X-rays, optical (Bond,
1997) and radio (Frail \etal 1997) followed the $\gamma$-rays. The
spectrum of the optical transient revealed a set of absorption lines
associated with Fe II and Mg II with a redshift $z=0.835$ (Metzger
\etal 1997) demonstrating the cosmological origin of GRBs. Radio
emission was observed first one week after the burst (Frail \etal
1997). This emission showed intensive oscillations which were
interpreted as scintillation (Goodman 1997). The subsequent
disappearance of these oscillations after about three weeks enabled
Frail \etall (Frail \etal 1997) to estimate the size of the fireball
at this stage to be $\sim 10^{17}$cm.  The observation that the radio
emission was initially optically thick (Frail \etal 1997), yielded a
similar estimate to the size (Katz \& Piran 1997). This size
immediately implies that the afterglow is expanding relativistically!

A dozen GRB afterglows have been discovered so far. It will be
impossible to discuss all those here. Worth mentioning are however,
GRB971214 and GRB980425. GRB971214 was a rather strong burst.  A
redshift of 3.42 was measured for the galaxy that is at the position
of GRB971214 (Kulkarni \etal 1998). For isotropic emission this large
redshift implies an energy release of $10^{53}$ergs \footnote{This
value is obtained for $\Omega=1$ and $H_0=65$km/sec/Mpc. The familiar
value of $3 \times 10^{53}$ (Kulkarni \etal 1998) is obtained for
$\Omega = 0.3$ and $H_0=0.55$km/sec/Mpc} in $\gamma$-rays alone.

GRB980425 was a moderately weak burst with a peak flux of $3 \pm 0.3
\times 10^{-7}{ \rm ergs \ cm^{-2} \ sec^{-1}}$. It was a single peak
burst with a rise time of 5 seconds and a decay time of about 25
seconds. The burst was detected by BeppoSAX (as well as by BATSE)
whose WFC obtained a position with an error box of $8'$. Inspection of
an image of this error box taken by the New Technology Telescope (NTT)
revealed a type Ic supernova SN1998bw that took place more or less
at the same time as the GRB (Galama \etal 1998b).  Since the
probability for a chance association of the SN and the GRB is only
$1.1 \times 10^{-4}$ it is likely that this association is real. The
host galaxy of this supernova (ESO 184-G82) has a redshift of
$z=0.0085 \pm0.0002$ putting it at a distance of $38 \pm 1$Mpc for
H=67km/sec Mpc. The corresponding $\gamma$-ray energy is $5 \times
10^{47}$ergs.

\section{Afterglow Models}

Afterglow observations provide a wealth of data in different
wavelengths and over a period of days, weeks and months. This should
be compared with the brief few second emission of the GRB. At the same
time modeling the afterglow is much simpler than modeling the
GRB. Consequently many efforts were devoted during the last year to
this problem.

\subsection{General Considerations}

According to the general picture afterglow is produces by shock
accelerated particles when the relativistic shell encounters the
surrounding ISM.  We can estimate the conditions of the shock
accelerated particles if we know the Lorentz factor of the ejecta,
$\gamma$, and the density of the ISM, $n_{ism}$.  Using the
relativistic shock conditions and assuming equipartition between the
different energy channels (protons energy, electrons energy and
magnetic field energy) we can estimate the typical electron's energy
$\ga_e \sim \epsilon_e(m_p/m_e) \ga$ and the typical magnetic field:
$B \sim \epsilon_B^{1/2} \ga \sqrt {m_p c^2 n_{ism}}$ ($\epsilon_e$
and $\epsilon_B$ are parameters of order unity). Given $\ga_e$ and $B$
and assuming that the electrons energy distribution follows a power
law: $N(\ga_e) \propto \ga_e^{-p}$ one can calculate the resulting
synchrotron radiation spectrum. These calculations has been quite
successful as can be seen, for example, in the observational fit of
Galama \etall (Galama \etal 1998a) to the theoretical spectrum of Sari
\etall (Sari, Piran \& Narayan 1998).

Estimating the light curve requires the knowledge of how the
conditions at the shock change with the observer time. Since those
depend on $\ga$ we need to know $\ga(t_{obs})$.  The $\ga(R)$ relation
is given by the hydrodynamics of the ejecta.  Two extreme limits of
this relation arise: $\ga \propto R^{-3/2}$ for an adiabatic expansion
(Blandford \& McKee 1976), and $\ga \propto R^{-6}$ for a radiative
expansion (when all the energy generated by the shock is radiated
away) (Cohen, Piran \& Sari 1998) \footnote{Note that this power is different from the commonly assumed
$\ga \propto R^{-3}$ for a radiative solution.}.  Emission of a
significant fraction but not all the energy would lead to intermediate
$\ga(R)$ relation. For a given model the $\ga(R)$ relation can be
substituted into the common equation for the observer time: $
t_{obs}=R/2 c \ga^2 $ to yield $\ga(t_{obs})$ \footnote{Care should be
exercised when using this last equation as strictly speaking it is
valid only for a shell moving at a constant velocity and it considers
only radiation that emerges along the line of sight.  Taking into
consideration the fact that in the realistic situation the shell is
decelerating and that one detects photons that are not just from the
line of sight one obtains a similar relation but the factor 2 is
replaced by another constant (which varies between 4 and 10) depending
on the specific model (Sari 1998, Waxman 1997, Panaitescu \& M\'esz\'aros 1998).}.  For $\ga
\propto R^{-n}$ we have $\ga \propto t_{obs}^{-1/(2+1/n)}$.  Thus $\ga
\propto t_{obs}^{-3/8}$ for adiabatic expansion and $\ga \propto
t_{obs}^{-6/13}$ for radiative expansion. Note that if $R$ is
practically constant (as would be the case during the sideway
expansion of a jet - discussed later) this power law will change only
slightly to $\ga \propto t_{obs}^{-1/2}$. The number of emitting
electrons behaves like $R^3$ and varies like $\ga^{-3/n} \propto
t_{obs}^{3/(2n+1)}$.

\subsection{Afterglow Transitions}

Both light curves of GRB970228 and GRB970508 shows a single unbroken
power law for as long as the afterglow could be detected. It is
interesting to compare this fact with afterglow models.

Several transitions should occur during the expansion of the ejecta
and the afterglow emission. At first the expansion should be radiative
as the shock accelerated particles cool rapidly compared with the
hydrodynamic time scale. As the shell slows down the particles become
less energetic, the cooling time increases and the evolution becomes
adiabatic.  The expansion at this stage becomes self-similar and it is
described by the Blandford-McKee (Blandford \& McKee 1976) solution.  A second
transition takes place when enough external mass has been accumulated
and the shell becomes Newtonian with $\ga \sim 1$. At this stage the
solution switches to the well known Sedov-Taylor solution. For
adiabatic evolution with no energy losses this transition should take
place at the Sedov length $R \sim l$.  For a spherical shell with
$10^{52}$ergs and $n_{ism}=1{\rm cm}^{-3}$, we find $ l \sim 2 \times
10^{18}$cm corresponding to a transition around two years after the
GRB. Any radiation losses shorten the time for this transition.

A third transition from quasi-spherical to non-spherical expansion
occurs if the ejecta is non-spherical. For an ejecta with an opening
angle $\theta$ this transition take place when $\ga \sim
\theta^{-1}$. For $\ga> \theta^{-1}$ the shell  behaves as if it is
a part of a spherical shell. For $\ga < \theta^{-1}$ the non-spherical
behaviour dominates, the jet expands rapidly sideways collecting more
and more ISM and slowing down rapidly with $\ga(R) \sim \exp
[-R/l\theta^{2/3}]$ (Rhoads 1998). In this non-spherical expansion
regime the radiation which was earlier beamed with an opening angle
$\theta$, is beamed into a cone with an opening angle $\ga^{-1}$. This
leads to a strong decreases of the observed flux as a function of
observed time. The solid angle into which the radiation is beamed
increases like $\ga^{-2} \propto t_{obs}$. Therefore the observed flux
decreases by approximately one power of $t_{obs}$ relative to a
corresponding quasi-spherical expansion.  Assuming, again, adiabatic
expansion the transition to non-spherical expansion takes place at
$t_{obs} = (l/2c) \theta^{10/3}$, corresponding for canonical
parameters and for $\theta =0.1 \sim 6^o$ to less than one day after
the GRB.

The lack of breaks in the observed light curves suggest that we have
not seen any transitions. Is this consistent with the theory? The
transition from radiative - adiabatic transition take place quite
early after the GRB.  Furthermore it does not have a strong effect on
the light curve.  It could have easily been missed due to lack of
early observations, or because the data is not accurate enough to show
it.  The Newtonian transition is rather late - a year or so after the
burst. In most cases the afterglow would be too weak to be detected  so
late\footnote{Recall that relativistic effects enhance strongly the
observed radiation.  During the Newtonian phase the afterglow is not
much stronger than a usual SNR}.

The regular power law behaviour of the optical afterglow of GRB970228
and GRB970508 suggests that there was no significant beaming in these
two events.  The optical light curve of GRB970508 shows a rapid rise
during the first two days and only then the power law decline
begins. It has been suggested that GRB970508 was beamed and we were
outside the initial beam. The rise corresponds to the increase in the
observed emission as the beam of this afterglow broadened after the
transition from quasi - spherical to non - spherical expansion. There
are two problems with this interpretation. First the decay of the
optical afterglow like $t^{-1.2}$ fits well a spherical or quasi -
spherical expansion and it does not fit the much faster decline of the
non - spherical phase of a jet. Second, it is not clear how was the
GRB detected in the first place if we were outside the viewing angle
of the original jet.

\section{New Puzzles}

Afterglow observations fit well the fireball picture that was
developed for explaining the GRB phenomena. The available data is not
good enough to distinguish between different specific models. But in
the future we expect to be able to distinguish between those models
and even to be able to determine the parameters of the burst (like $E$
and $\ga_0$ if the data is taken early enough), the surrounding ISM
density and the intrinsic parameters of the relativistic shock
$\epsilon_e$, $\epsilon_B$ and $p$.  Still the current data is
sufficient to raise new puzzles and present us with new questions.

\begin{itemize}
\item{\bf Why  afterglow accompany some GRBs and not others?}

X-ray, Optical and radio afterglows have been observed in some bursts
but not in others. According to the current model afterglow is
produces when the ejecta that produced the GRB is shocked by the
surrounding matter.  Possible explanations to this puzzle invoke
environmental effects.  A detectable afterglow might be generated
efficiently  in some range of ISM densities and inefficiently in
another.  High ISM densities would slow down of the ejecta more
rapidly. This could make some afterglows detectable and others
undetectable.  ISM absorption is another alternative. While most
interstellar environments are optically thin to gamma-rays high
density ISM regions can absorb and attenuate efficiently x-rays and
optical radiation.

\item {\bf Jets and the Energy of GRB971214}

How can we explain the $ 10^{53}$ergs required for isotropic emission
in GRB971214?  This amount is larger than what all current models can
produce.  This problem can be resolved if we invoke beaming, with
$\theta \sim 0.1$.  However, such beaming is ruled out in other
afterglows for which there are good data. It would have been much
simpler if a possible (but unlikely) interpretation of the observed
spectrum to have a redshift of 0.444 (Kulkarni \etal 1998) could be
adopted.

\item{\bf GRB980425 and SN1998bw}

SN1998bw (and the associated GRB980425) is a factor of a hundred
nearer than a typical GRB (which are expected to be at $z \sim
1$). The corresponding (isotropic) gamma-ray energy, $\sim 5 \times
10^{47}$ergs, is four order of magnitude lower than a regular
burst. This can be in agreement with the peak flux distribution only
if the bursts with such a low luminosity compose a very small fraction
of GRBs. This leads naturally to the question is there an
observational coincidence between GRBs and SNs?  To which there are
conflicting answers (Wang \& Wheeler 1998, Kippen \etal 1998, Bloom
\etal 1998).

\end{itemize}
\section{The Inner Engine - What Powers GRBs}

As have been stressed earlier we cannot observe the ``inner engine'' that
powers a GRB. The observed variability time scale implies that this
source must be compact - smaller than $10^3$km. Even if such a source
would have produced gamma-rays it would have been optically thick and
undetectable.  The source produces a relativistic particle flow and
the observed radiation is produced by this flow far away from the
center.  This is not an unfamiliar situation in astronomy. The sun's
core could not be observed directly until solar neutrino experiments
began. As the source is not observed directly we can infer on its
nature only indirectly:

\begin{itemize}

\item {\bf (i) Energetics:} The source  should produce the needed energy  -
$\sim 10^{52}$ergs for isotropic emission, $\theta^2 /4 \pi$ times
that for a jet with an opening angle $\theta$. It should also be
capable of producing (or rare occasions?) the observed $10^{53}$ergs
required for events like GRB971214.

\item{\bf (ii) Relativistic Flow:} The source must produce
a relativistic particle flow. This requires that there will be a small
(but not too small) baryonic load: $ m \sim 5 \times 10^{-5} m_\odot
(E/10^{52}{\rm ergs})(\ga/100)^{-1}$.

\item{\bf (iii) Duration:}
According to the internal shocks scenario the duration of the burst is
$\Delta/c$ which in turn equals to the time that the inner engine is
active.

\item{\bf (iv) Variability:}
The observed variability implies that the source should be compact
with $R_{source} \le 10^3{\rm km} (\delta t /0.003{\rm sec})$. The
combination of the last two items rules out ``explosive'' sources that
produce a single pulse with $T \sim \delta t \sim R_{source}/c$.

\item{\bf (v) Rate:}
GRBs take place at a rate of $10^{-5}-10^{-6} /{\rm year \ galaxy}$.
beaming will increase this estimate by $ 4 \pi /\theta^2$
\end{itemize}

Current models for the internal engine include: (1) Binary neutron
star merger (Eichler, Livio, Piran, \& Schramm 1989; Paczy\'nski
1986), (2) Failed Supernova (Woosley 1993) or hypernova (Paczy\'nski
1997), and (3) Magnetic white dwarf collapse (Usov 1992; Duncan \&
Thompson 1992). All these model can produce, in principle
$10^{52}$ergs. In none of these models it is clear how is this energy
channeled to the essential relativistic flow. A black hole forms in
(1) and (2) and the GRB is powered by an accretion disk that forms
around it (Narayan, Paczy\'nski \& Piran 1992; Popham, Woosley \&
Fryer 1998). Narayan \etall (1992) suggested that the relativistic
flow is produced via magnetic field recombination in the disk. Katz
(1997) suggested that there is a pulsar like mechanism.  The dynamical
time of such an accretion disk is a few milliseconds.  Accretion takes
place on a viscous time, which is orders of magnitude larger.  Thus
the conditions concerning the duration and variability could in
principle be satisfied. However, at present it is impossible to
calculate from first principles how is this done (see however Popham,
Woosley \& Fryer 1998).

In the magnetic white dwarf collapse the relativistic energy flow is
carried by Poynting flux and not by particles. Here the energy source
is the magnetic field and the rotational energy of the  magnetic neutron star. Different considerations determine the duration of the activity.

The energy condition (i) and the variability (or size) condition (iv)
are satisfied by all three model.  It is possible, but not calculable,
that conditions (iii) concerning the overall duration is satisfied. In
all three models the question how is the relativistic flow generated
(condition (ii)) remains open\footnote{Paczy\'nski (Paczy\'nski 1997) in the
hypernova model suggests that the relativistic flow is produced when
an initially Newtonian shock is accelerating while interacting with
external matter with a decreasing density profile.}.

The last condition concerning the rate is satisfied by the binary
neutron star merger model (Piran 1992; Piran, Narayan \& Shemi
1992), which is the only model
based on an independently observed phenomenon. This agreement holds if
the burst is more or less isotropic. Significant beaming will, of
course, change this. Binary neutron star mergers produce a unique
specific gravitational radiation signal that cannot be
misinterpreted. They are the best candidates for sources of
gravitational radiation signals.  A clear prediction of this model is
that such a signal should appear in coincidence with a GRB
(Kochaneck \& Piran 1993). Thus, a unique feature of this model is that
it can be tested and verified in the nearby future when 
gravitational detectors LIGO and VIRGO will become operational.


\section{Concluding Remarks}

GRB astronomy has undergone two major revolutions during the last
decade. The first have shown that GRBs are cosmological, the second
have confirmed the basic features of the prevailing fireball model.

Both revolutions were observationally driven by new satellites. Still
their basic findings have been predicted earlier on by theoretical
studies. The isotropy of the pre-BATSE GRB sample has motivated
several suggestions that GRBs are cosmological (Van Den Bergh 1983;
Paczy\'nski 1986; Hartman \& Blumenthal 1989). On the other hand
analysis of the neutron star merger phenomenon (Eichler, Livio, Piran,
\& Schramm 1989) has lead to the suggestion that it would generate a
GRB. This has given another support to possibility of a cosmological
origin.

Relativistic motion was then suggested to overcome the compactness
problem. This problem posed a serious objection to cosmological  (and hence
very luminous) GRB sources. This has lead to the fireball model, whose
general features, and in particular relativistic motion, were
confirmed by  afterglow observations.

In spite of all this progress the ``inner engine'' that powers GRBs is not
understood yet. It remains hidden. All that we have at present is only
indirect circumstantial evidence on its behaviour. The origin of GRBs
is still a puzzle.

I would like to thank Ehud Cohen, Jonathan Granot, Jonathan Katz,
Ramesh Narayan and Re'em Sari for many helpful discussion. This
research was supported by a US-Israel BSF grant 95-328 and by a NASA
grant NAG5-3516.

\section{References}

\vspace{1pc}
\re
1.    Blandford, R. D., \& McKee, C. F., 1976, {\it Phys. of Fluids} {\bf 19}, 1130.
\re
2.    Bloom, J. S.,  \etall 1998, astro-ph/9807050.
\re
3.    Bond, H. E., 1997, IAU circ.  6665.
\re
4.    Cohen, E., \&  Piran, T., 1995, {\it Ap. J.} {\bf 444}, L25.
\re
5.    Cohen, E., \etal, 1997, {\it Ap. J.} {\bf 480}, 330.
\re
6.    Cohen, E., Piran, T., \& Sari, R.,  1998, {\it Ap. J.} {\bf 509} in press,
astro-ph/9803258.
\re
7.    Costa, E., \etal, 1997, {\it Nature} {\bf 387}, 783.
\re
8.    Duncan, R., \& Thompson, C., 1992, {\it Ap. J. Lett.} {\bf 392}, L9.
\re
9.    Eichler, D., Livio, M., Piran, T., \& Schramm, D. N. 1989,
{\it Nature} {\bf 340}, 126.
\re
10.    Fenimore E. E. \etal, 1993, {\it Nature} {\bf 366}, 40.
\re
11.   Frail, D. A., \etal, 1997, {\it Nature} {\bf 389}, 261.
\re
12.   Galama, T. J., \etal, 1997, {\it Nature} {\bf 387}, 497.
\re
13.   Galama, T. J., \etal, 1998a, {\it Ap. J.} in press.
\re
14.   Galama, T. J., \etal, 1998b, submitted to {\it Nature}, astro-ph/9806175.
\re
15.   Goodman, J., 1986, {\it Ap. J.} {\bf308}, L47.
\re
16.   Goodman, J., 1997, {\it New Astronomy} {\bf 2}, 449.
\re
17.   Hartmann, D. H., \& Blumenthal , G. R., 1989, {\it Ap. J.} {\bf 342}, 521.
\re
18.   Katz, J. I., 1994, {\it Ap. J.} {\bf 422}, 248.
\re
19.   Katz, J. I., 1997, {\it Ap. J.} {\bf 490}, 633.
\re
20.   Katz, J. I., \& Piran, T., 1997, {\it Ap. J.} {\bf  490}, 772.
\re
21.   Kippen, R. M., \etall 1998, astro-ph/9806364.
\re
22.   Kobayashi, S., Piran, T., \&  Sari, R., 1997, {\it Ap. J.} {\bf 490}, 92.
\re
23.   Kochaneck C., \&  Piran, T., 1993, {\it Ap. J. Lett.} {\bf 417}, L17.
\re
24.   Krolik, J. H., \& Pier, E. A., 1991, {\it Ap. J.} {\bf 373}, 277.
\re
25.   Kulkarni, S., \etall 1998, {\it Nature} {\bf 393}, 35.
\re
26.   Meegan,  C. A., \etal, 1992, {\it Nature} {\bf 355}, 143.
\re
27.   M\'esz\'aros, P., \& Rees, M. J., 1992, {\it MNRAS} {\bf 258}, 41P.
\re
28.   M\'esz\'aros, P., \& Rees, M. J., 1997,  {\it Ap. J.} {\bf 476}, 232.
\re
29.   Metzger, M. R., \etal, 1997, {\it Nature} {\bf 387}, 878.
\re
30.   Mochkovitch, R., Maitia, V., \& Marques, R., 1995,
in: {\it Towards the Source of Gamma-Ray Bursts, Proceedings of 29th
ESLAB Symposium},  Bennett, K., \& Winkler, C., Eds., 531.
\re
31.   Narayan, R., Paczy\'nski, B., \& Piran, T., 1992, {\it Ap. J.
Lett.} {\bf 395}, L83.
\re
32.   Paczy\'nski, B., 1986, {\it Ap. J.} {\bf 308}, L51.
\re
33.   Paczy\'nski, B., 1990, {\it Ap. J.} {\bf 363}, 218.
\re
34.   Paczy\'nski, B. \& Rhoads, J., 1993, {\it Ap. J. Lett.} {\bf 418}, L5.
\re
35.   Paczy\'nski, B., \& Xu, G.,  1994, {\it Ap. J.} {\bf 427}, 709.
\re
36.   Paczy\'nski, B., 1997, {\it Ap. J. Lett.} {\bf 484}, L45.
\re
37.   Panaitescu, A., \& M\'esz\'aros, P., 1998, {\it Ap. J.} {\bf 492}, 683.
\re
38.   Piran, T., 1992, {\it Ap. J. Lett.} {\bf 389}, L45.
\re
39.   Piran, T., Narayan, R., and Shemi, A., 1992, in: AIP
Conference Proceedings {\bf 265}, {\it Gamma-Ray Bursts, Huntsville,
Alabama, 1991}, Paciesas, W.S., and Fishman, G.J., Eds. (New York:
AIP), p. 149.
\re
40.   Piran, T., 1997, in: {\it Some Unsolved Problems in
Astrophysics}, Bahcall, J. N., and Ostriker, J. P., Eds.,
Princeton University Press.
\re
41.   Popham, R., Woosley, S., E., \& Fryer, C., 1998, astro-ph/9807028.
\re
42.   Rees, M. J., \& M\'esz\'aros, P. 1994, {\it Ap. J. Lett.} {\bf 430}, L93.
\re
43.   Rhoads, J. E., 1998, preprint.
\re
44.   Sari, R., \& Piran, T., 1997, {\it Ap. J.} {\bf 485}, 270.
\re
45.   Sari, R., 1998, {\it Ap. J. Lett.} {\bf 494}, L49.
\re
46.   Sari, R., Piran, T., \& Narayan, R., 1998, {\it Ap. J. Lett.} {\bf 497}, L41.
\re
47.   Shemi, A., \& Piran, T., 1990, {Ap. J.} {\bf 365}, L55.
\re
48.   Usov, V. V., 1992, {\it Nature} {\bf 357}, 472.
\re
49.   Van Den Bergh, S., 1983, {\it Astrophys. Space Sci.} {\bf 97}, 385.
\re
50.   van Paradijs, J., \etal, 1997, {\it Nature} {\bf 386}, 686.
\re
51.   Wang, L., \& Wheeler, J. C., 1998, {\it Ap. J. Lett.} in press, Astro-ph/9806212.
\re
52.   Waxman, E., 1997, {\it Ap. J. Lett.} {\bf 491}, L19.
\re
53.   Wijers, A. M. J., Rees, M. J., \& M\'esz\'aros, P., 1997, {\it MNRAS}
{\bf 288}, L51.
\re
54.  Woosley, S. E. 1993, {\it Ap. J.} {\bf 405}, 273.

\vspace{1pc}

\end{document}